\documentclass[12pt,a4]{article}
\usepackage{graphicx,amsfonts}
\usepackage{longtable}
\usepackage{amsmath}
\usepackage[all]{xy}

\newcommand{\be}{\begin{equation}}
\newcommand{\ee}{\end{equation}}
\newcommand{\ba}{\begin{eqnarray}}
\newcommand{\ea}{\end{eqnarray}}
\newcommand{\baa}{\begin{eqnarray*}}
\newcommand{\eaa}{\end{eqnarray*}}

\begin{document}

\title{Effects of decoherence on the shot noise in carbon nanotubes}
\author{Cristina Bena\\
{\small \it Laboratoire de Physique des Solides, Universit\'e Paris-Sud},
\vspace{-.1in}\\{\small \it  B\^at.~510, 91405 Orsay, France}\\
{\small \it Institute de Physique Th\'eorique, CEA/Saclay},
\vspace{-.1in}\\{\small \it  Orme des Merisiers, 91190 Gif-sur-Yvette Cedex, France}}

\maketitle

\begin{abstract}
We study the zero frequency noise in an interacting quantum wire connected to leads,
in the presence of an impurity. In the absence of quasiparticle decoherence
the zero-frequency noise is that of a non-interacting wire.
However, if the collective, fractionally-charged modes have a finite lifetime, we find that the zero-frequency noise may still exhibit signatures of charge fractionalization, such as a small but detectable reduction of the ratio between the noise and the backscattered current (Fano factor). We argue that this small reduction of the Fano factor is consistent with recent observations of a large reduction in the experimentally-inferred Fano factor in nanotubes (calculated  assuming that the backscattered current is the difference between the ideal current in a multiple-channel non-interacting wire and the measured current).
\end{abstract}

\noindent Shot noise has been long used to extract information about the nature of the elementary excitations in a system. Thus, shot noise measurements have been used extensively to study exotic phases, such as the ones arising in strongly interacting one-dimensional systems. For example they confirmed the presence of fractionally charged quasiparticles in a fractional quantum Hall fluid \cite{glattli}. While various proposals have been made to use shot noise to characterize other one-dimensional systems such as carbon nanotubes \cite{benaold,dolcini,n1,n2,n3}, a clear confirmation of the fractionalized nature of the elementary quasiparticles in nanotubes is still lacking. The major effect that obscures the fractional character of the excitations in nanotubes (and other non-chiral one-dimensional systems) is the presence of metallic leads: at small frequencies the noise probes the physics of the non-interacting leads, and not the physics of the nanotube. Thus, while in a chiral Luttinger liquid (LL) (such as a fractional quantum Hall effect (FQHE) edge state system), the Fano factor, defined as the ratio betwen the noise and the backscattered current, is equal to the fractional charge $g e$, in a nanotube the Fano factor has been predicted \cite{dolcini} to be simply equal to the electron charge $e$.

Recent experiments \cite{kontos,noise0} have measured the Fano factor in nanotubes, and observed nevertheless a reduced Fano factor. The value of the observed Fano factor in Ref.~\cite{kontos} is roughly $0.4 e$ -- too small to be explained by the non-interacting physics of the leads \cite{dolcini}, but also too large to be consistent with the presumed value of the fractional charge in nanotubes ($0.2 e$) \cite{benaold}.

We propose that this reduction of the Fano factor is the result of a combination of three factors: The first are the interactions in the wire. The second is the quasiparticle decoherence, which generates a finite lifetime for the fractionally charged quasiparticles; a possible cause for the appearance of a finite quasiparticle lifetime is the band curvature \cite{glazman,teber}, which is usually ignored when the standard bosonisation procedure is performed. The third factor comes from the experimental method of measuring the backscattered current in a non-chiral LL, for which the right-moving modes and the left-moving modes are not spatially separated. This backscattered current cannot be measured directly as in a fractional quantum Hall edge system, but it is assumed to be the difference between the current in the absence of an impurity (estimated using the ideal conductance of a nanotube $4 e^2/h$), and the measured current.

We claim that the combination of these three factors yields a reduced value of the Fano factor consistent with the experimental measuremens. Indeed, the combination of interactions and decoherence leads to a reduction in the conductance of the nanotube, even in the absence of impurities, from $4 e^2/h$ to $4 g_e e^2/h$, where $g_e$ is an effective ``fractional charge''-type parameter that is close but not equal to $1$. This generates large errors in the evaluation of the backscattered current (that use the {\it non-interacting} conductance $4 e^2/h$).  Moreover, the zero-frequency noise is also reduced from $e I_B$ to $g_e e I_B$. This yields a reduction of the ``true'' Fano factor of the system to $g_e e$ instead of $e$. We show that the combination between the reduction of the ``true'' Fano factor and that of the ideal conductance (which indirectly feeds into the estimation of the backscattered current) will yield a reduction of the ``inferred'' Fano factor of the system to the value observed in the experiments \cite{kontos}.
Our analysis also allows us to propose a direct experimental test to extract the value of $g_e$ from the dependence of the zero-temperature noise on the conductance of the wire.

\bigskip

\begin{figure}
\vspace{0.3cm}
\begin{center}
\includegraphics[width=8cm]{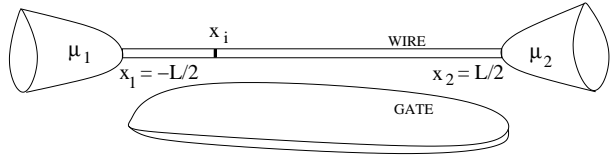}
\vspace{0.15in} \caption{\small  A quantum wire
with an impurity adiabatically coupled to metallic leads. The leads are held at different
chemical potentials $\mu_1$ and $\mu_2$.}
\label{setup}
\end{center}
\end{figure}

A quantum wire connected to metallic leads is presented in Fig.~\ref{setup}. The interaction parameter $g(x)$ is space-dependent and
its value is $g$ in the bulk of the wire, and 1 in the
leads. 

The ideal AC conductance for this system in the absence of impurity scattering, and in the absence of decoherence  is given by \cite{dolcini,safi}:
\be
G_0(\omega)=\frac{e^2}{h}\Big[1+\gamma \frac{2 i \sin(\omega/\omega_L)}{e^{-i \omega/\omega_L}-\gamma^2 e^{i \omega/\omega_L}}\Big]\,,
\label{g0}
\ee
where $\gamma=(1-g)/(1+g)$, $\omega_L=v_F/g L$, $v_F$ is the Fermi velocity, and $L$ is the length of the wire.
The effects of decoherence can at first naively be taken into account by adding an imaginary self energy to the propagating Green's function, or equivalently using the standard field theory procedure of adding an imaginary part to the energy such that $\omega \rightarrow \omega+i \delta$, where $\delta \propto 1/\tau$ is the inverse quasiparticle lifetime.

We focus mainly on the DC conductance, and take the limit $\omega\rightarrow 0$, while keeping $\delta$ finite. This yields:
\be
G_0=\frac{e^2}{h}\Big[1+\gamma \frac{2 \sinh(\delta/\omega_L)}{e^{\delta/\omega_L}-\gamma^2 e^{-\delta/\omega_L}}\Big]\,.
\label{g1}
\ee

We can study a few limits of this equation. For example, if the tube is much shorter than the coherence length, such that
$\delta \ll \omega_L$, the conductance of the wire goes back to the non-interacting value of $G_0=e^2/h$. In the opposite limit, when the tube is much longer than the coherence length we find a value for the conductance of $G_0=2 g/(1+g) e^2/h$, which is the conductance of an ideal semi-infinite wire - metal junction. In the intermediate regime, for example if the coherence length is a few times the length of the wire, the conductance is reduced to $G_0=g_e e^2/h$, with $g_e$ close to, but slightly smaller that $1$. We believe this situation is most relevant for the existing experiments. Also, for a nanotube which has four channels of conduction $G_0=4 g_e e^2/h$.

In the presence of an impurity, a portion of the current will be backscattered, such that $I=I_0-I_B$, with $I_0 \equiv G_0 V$.
In general, at low temperatures and for voltages larger than $\omega_L$, the backscattered current has a non-linear dependance on voltage \cite{dolcini,dolcini_current}, consistent with the LL theory. However, for voltages smaller than $\omega_L$, the dependence of the backscattered current on voltage is linear \cite{dolcini,dolcini_current}. The impurity-induced noise in the current is \cite{saleur}:
\be
S=2 g_e e I_B {\cal{T}}(E) \coth\Big(\frac{g_e e V}{2 k_B T}\Big)+4 k_B T G_0 {\cal{T}}^2(E)\,,
\label{noise}
\ee
where $ { {\cal{T}}(E) \equiv {1 \over G_0}  {dI \over d{\cal V}}|_{{\cal V}=E}} $ is the
energy-dependent transmission of the wire, $I$ is the transmitted current, and the backscattered current is
\be
I_{B}=G_0 V - I=G_0 V-\int_0^V dE \frac{dI}{d{\cal V}}\bigg|_{{\cal V}=E}\,.
\ee
For $g_e=1$ Eq.~(\ref{noise}) describes adequately the noise for a non-interacting system \cite{noise00}. For an infinite LL ($g_e=g$) Eq.~(\ref{noise}) correctly describes the noise if the backscattering is small (${\cal{T}}(E)\approx 1$), moreover this formula is accurate up to a few percent error even for lower transmission coefficients, up to a bare transmission of the barrier of $\approx 50\%$ \cite{saleur}. Eq.~(\ref{noise}) also describes accurately the noise of an interacting LL with leads \cite{dolcini,nagaosa}; for this system it was proposed that $g_e$ is equal to $1$. The arguments we presented above for the ideal conductance of a non-chiral LL connected to metallic leads imply that Eq.~(\ref{noise}) with a renormalized of $g_e \approx 0.9$ should be used to describe the noise in a nanotube.

We can now estimate the value of the Fano factor and compare it with the experimentally measured value.
The Fano factor is given by the ratio between the noise and the backscattered current. The noise is measured directly in experiments. However the backscattered current for a non-chiral LL cannot be measured directly but is inferred \cite{kontos} from the measured transmitted current, $I$, using:
\be
I^*_{B}=4\frac{e^2}{h} V- \int_0^V dE \frac{dI}{d{\cal V}}\bigg|_{{\cal V}=E}.
\ee
Thus, in the limit of small temperature, the experimentally-inferred Fano factor is given by:
\be
F^*=\frac{S}{2 e I_B^*}=\frac{g_e I_B {\cal{T}}(E)}{I_B^*}=g_e {\cal{T}}(E) \frac{4 g_e \frac{e^2}{h} V-\int_0^V dE \frac{dI}{d{\cal V}}\big|_{{\cal V}=E}}{4\frac{e^2}{h} V- \int_0^V dE \frac{dI}{d{\cal V}}\big|_{{\cal V}=E}}
\ee

One should note that the real Fano factor, $S/2 e I_B=g_e  {\cal{T}}(E)$, would be accessible if one could measure directly the backscattered current. In the limit of good transmission, ${\cal{T}}(E)\approx 1$, the real Fano factor would yield the value of the fractional charge $g_e$. However, given the experimental difficulties to isolate the backscattered current, one estimates instead $F^*$. It is not hard to see how the value of $F^*$ may easily differ from the non-interacting value $1$, even if $g_e$ is very close to $1$. For example, assuming that $I=\int_0^V dE dI/d{\cal V}|_{{\cal V}=E}=\alpha 4 g_e e^2/h V$, ($\alpha$ is always smaller than $1$), in the limit of very good transmission ${\cal{T}}(E)\approx 1$, we get
\be
F^* \approx g_e^2 \frac{1-\alpha}{1-g_e \alpha}\,.
\ee
Thus, for $\alpha=0.9$, and $g_e=0.9$, $F^*\approx 0.4$, in very good agreement with the experimental observations. Note that this value is reduced further by a factor of ${\cal{T}}(E)$ in the regime where the transmission deviates significantly from unity.

An important observation is that the measured Fano factor $F^*$ depends strongly on the transmission of the wire $\alpha$ (decreasing towards zero with increasing $\alpha$ towards a perfect transmission). On the other hand the real Fano factor is roughly independent or increases weakly with increasing the differential conductance. Moreover, in the regime where the dependence on the current on voltage is non-linear, $\alpha$ will also depend on voltage, and the measured Fano factor $F^*$ should also depend on the voltage.

We also propose a direct test of our hypothesis, and a direct way of measuring $g_e$ by analyzing the dependence of the zero-temperature noise in units of $2 e V dI/dV$ on $I/V$. Rewriting Eq.(\ref{noise}) as:
\be
S=2 g_e e (V-I/G_0) dI/dV\,,
\ee
we can see that
\be
\tilde{S}=S/(2 e V dI/dV)=g_e [1-1/G_0 I/V]\,.
\ee
Thus one expects $\tilde{S}$ to have a linear dependence on $I/V$, with a $g_e/G_0=h/4 e^2$ slope, but with an $x$-axis intercept at $x_0=G_0=g_e 4 e^2/h$. The experimental data presented in Ref~\cite{kontos} depict only $S/I$ as a function of $dI/dV$, which should have a slightly different behavior, but for which we can already observe a small deviation from $1$ in the intercept of the fit to the data with the $x$-axis. The data presented in Ref.~\cite{kontos} is however hard to interpret due to both finite temperature effects, as well as non-linearities in ${\cal{T}}(E)$  (due to the scattering of quasiparticles at the two contacts). More precise data at lower temperature is needed for a detailed quantitative comparison with the theoretical predictions.

\bigskip

In conclusion, we have found that the interplay between decoherence and interactions in non-chiral LL's can explain the recent experimental observation of Ref.\cite{kontos}, that finds a substantial reduction of the Fano factor from the non-interacting value. Indeed, it has been predicted that the Fano factor of a Luttinger liquid connected with metallic leads should be the same as that of a system without interactions. Our analysis has revealed that in the presence of interactions and decoherence the {\it true} Fano factor of the wire is reduced to a value $g_e$ that is close, but not equal to $1$.

Furthermore, in an experiment one does not have access to the {\it true} Fano factor, but to the {\it inferred} Fano factor, which is computed using a value of the backscattered current that is not measured directly, but estimated indirectly from the measured conductance using the non-interacting ideal conductance. We have demonstrated that even small reductions to the {\it true} Fano factor and to the ideal conductance dramatically lower the {\it inferred} Fano factor and can explain the value observed in the experiments presented in Ref.\cite{kontos}.

Last, but not least, we have proposed a direct test for extracting the value of the renormalized ``fractional charge'' parameter $g_e$ from the dependence of the zero-temperature noise (renormalized by the voltage and the differential conductance) on $I/V$. A more comprehensive study of the Fano factor in the presence of interactions and decoherence, as well as an analysis of the interplay between various sources of decoherence is under way.

\bigskip

{\noindent {\bf Aknowledgements} \\I would like to thank H.~Bouchiat, R.~Deblock, C.~Glattli, T.~Kontos, I.~Safi and P. Simon for helpful discussions. This work was partially supported by a Marie Curie Action under FP6.}

\end{document}